\documentclass[prl,showpacs,tightenlines,twocolumn,preprint,10pt]{revtex4}
\usepackage{times}
\usepackage[T1]{fontenc}
\usepackage[latin1]{inputenc}
\usepackage{amsmath}
\usepackage{graphicx}

\begin{document}


\title{Quantum tetrahedral mean field theory of the magnetic susceptibility for the
pyrochlore lattice}

\author{A. J. García--Adeva}

\email{garcia@landau.physics.wisc.edu}

\author{D. L. Huber}

\affiliation{Department of Physics, University of Wisconsin--Madison, Madison WI 53706}

\begin{abstract}
A quantum mean field theory of the pyrochlore lattice is presented. The starting
point is not the individual magnetic ions, as in the usual Curie--Weiss mean
field theory, but a set of interacting corner sharing tetrahedra. We check the
consistency of the model against magnetic susceptibility data, and find a good
agreement between the theoretical predictions and the experimental data. Implications
of the model and future extensions are also discussed.
\end{abstract}

\maketitle

\textit{Introduction.-} Geometrically frustrated antiferromagnets with a pyrochlore
lattice exhibit a rich phenomenology which has received a vast amount of attention
during the last decade \cite{hfm.2000,ramirez.1994}. In the pyrochlore lattice,
the magnetic ions occupy the corners of a 3D arrangement of corner sharing tetrahedra.
Materials that crystallize in this structure exhibit anomalous magnetic properties
\cite{raju.1998, raju.1999, huber.2000}: The magnetic susceptibility follows
the Curie--Weiss law down to temperatures well below the Curie temperature.
Some of them exhibit long range order at very low temperatures, whereas others
behave as spin glasses, even though the lattice is almost perfect. There are
other compounds of this class that show short range order, and are regarded
as spin liquids.

In spite of the intensive research in these systems during the last decade,
there is still a lack of critical comparison between theory and experiment.
Even though there are a number of models and classical Monte Carlo simulations
\cite{moessner.1999} which qualitatively describe some of the experimental
results, some of the features found in the experimental data cannot be explained
by means of a classical theory, as are, for example, the maxima appearing in
the magnetic susceptibility at very low temperatures, which a classical model
cannot explain. However, there have been few attempts to investigate these systems
quantum mechanically. Harris and co--workers have studied the quantum \( s=\frac{1}{2} \)
Heisenberg antiferromagnet \cite{harris.1991}. Canals and Lacroix \cite{canals.1998},
have applied a perturbative approach to the density operator of a small cluster,
and found that the ground state is a quantum spin liquid.

In this work, we undergo the task of making such a quantum theory of the pyrochlore
lattice in the framework of the mean field theory. The goal of this work is
twofold: In one hand, we introduce a fully quantum mechanical mean field theory
of the Heisenberg antiferromagnet in the pyrochlore lattice for arbitrary spin
\( s \). In the other hand, we try to make this model as simple as possible,
in order to make it easy to compare it with experimental data and extract information
about the various interactions that play a role in these systems. Only the main
results of the model will be presented here, as a more detailed presentation
of the model will be published elsewhere. The starting point to build this MF
is not a set of interacting spins, but a set of coupled tetrahedra, which goes
back to the constant coupling approximation of Kastelejein and van Kranendonk
\cite{kastel.1956}. The magnetic susceptibility of this system is calculated
for both the interacting and non--interacting tetrahedra cases. We find that
the independent tetrahedra model fails to explain the behavior of experimental
data, and we justify this fact, whereas the MF theory describes quasi--quantitatively
the experimental data for a variety of systems.

\textit{The model.-} The Hamiltonian of the quantum Heisenberg model with nearest
neighbors in the presence of an applied magnetic field \( H_{0} \) in the pyrochlore
lattice can be put as \cite{smart.1966} 
\begin{equation}
H=-2J_{1}\sum _{\left\langle i,j\right\rangle }\vec{s}_{i}\cdot \vec{s}_{j}-H_{0}\sum _{i}s_{z_{i}},
\end{equation}
 where \( J_{1} \) is the negative exchange coupling, and \( \vec{s}_{i} \)
is the spin operator of the \( i \)--th magnetic ion. The Zeeman term has been
quoted in units of the Bohr magneton times the gyromagnetic ratio, so \( H_{0} \)
has dimensions of energy.

We start by considering the magnetic susceptibility of one tetrahedron. In this
simple case, the Hamiltonian can be easily diagonalized in terms of the total
spin representation of the tetrahedron, and the magnetization can be expressed
as 
\begin{equation}
\label{magnetization}
M=\frac{\sum _{S}g(S)Se^{j_{1}S(S+1)}Q_{S}(x)}{\sum _{S}g(S)e^{j_{1}S(S+1)}\frac{\sinh \left( \frac{2S+1}{2S}x\right) }{\sinh \left( \frac{x}{2S}\right) }},
\end{equation}
 where \( S \) represents the modulus of the total spin operator of the tetrahedron;
\( g(S) \) is the degeneracy associated to the total spin value \( S \), which
can be calculated by using Van Vleck's formula \cite{vanvleck.1959}, and are
listed in table \ref{table.degeneracies} for the values of the individual spins
\( s \) considered in this work; 
\begin{table}
{\centering \begin{tabular}{c|r|r|r|r|r|r|r|r|r|r|r|r|r|r|r|}
\cline{2-16} 
\multicolumn{1}{c|}{}&
\multicolumn{15}{|c|}{\( S \) }\\
\cline{2-2} \cline{3-3} \cline{4-4} \cline{5-5} \cline{6-6} \cline{7-7} \cline{8-8} \cline{9-9} \cline{10-10} \cline{11-11} \cline{12-12} \cline{13-13} \cline{14-14} \cline{15-15} \cline{16-16} 
\multicolumn{1}{c|}{}&
\multicolumn{1}{|c|}{0}&
\multicolumn{1}{|c|}{1}&
\multicolumn{1}{|c|}{2}&
\multicolumn{1}{|c|}{3}&
\multicolumn{1}{|c|}{4}&
\multicolumn{1}{|c|}{5}&
\multicolumn{1}{|c|}{6}&
\multicolumn{1}{|c|}{7}&
\multicolumn{1}{|c|}{8}&
\multicolumn{1}{|c|}{9}&
\multicolumn{1}{|c|}{10}&
\multicolumn{1}{|c|}{11}&
\multicolumn{1}{|c|}{12}&
\multicolumn{1}{|c|}{13}&
\multicolumn{1}{c|}{14}\\
\hline 
\hline 
\( s=1 \)&
3&
6&
6&
3&
1&
&
&
&
&
&
&
&
&
&
\\
\hline 
\( s=\frac{3}{2} \)&
4&
9&
11&
10&
6&
3&
1&
&
&
&
&
&
&
&
\\
\hline 
\( s=\frac{5}{2} \)&
6&
15&
21&
24&
21&
15&
10&
6&
3&
1&
&
&
&
&
\\
\hline 
\( s=\frac{7}{2} \)&
8&
21&
31&
38&
42&
43&
41&
36&
28&
21&
15&
10&
6&
3&
1\\
\hline 
\end{tabular}\par}
\caption{\label{table.degeneracies}Values of \protect\( g(S)\protect \) for the individual
spins, \protect\( s\protect \), considered in this work.}
\end{table}
 \( x=\frac{H_{0}S}{T} \), and \( j_{1}=\frac{J_{1}}{T} \) (the energies are
quoted in units of the Boltzmann constant, so they have dimension of absolute
temperature); and 
\begin{equation}
Q_{S}(x)=\frac{2S+1}{2S}\frac{\cosh \left( \frac{2S+1}{2S}x\right) }{\sinh \left( \frac{x}{2S}\right) }-\frac{1}{2S}\frac{\sinh \left( \frac{2S+1}{2S}x\right) }{\sinh \left( \frac{x}{2S}\right) }\coth \left( \frac{x}{2S}\right) .
\end{equation}

In the limit \( x\ll 1 \), we can define the magnetic susceptibility of the
tetrahedron as 
\begin{equation}
\label{suscep.tetra}
\chi ^{tet}=\frac{M}{H_{0}}=\frac{1}{T}\frac{\sum _{S}g(S)S(S+1)(2S+1)e^{j_{1}S(S+1)}}{\sum _{S}g(S)(2S+1)e^{j_{1}S(S+1)}}.
\end{equation}

The susceptibility per ion in the tetrahedron will be simply given by \( \hat{\chi }^{tet}=\frac{\chi ^{tet}}{4} \).

It is interesting to note that for \( j_{1}\ll 1 \), the susceptibility can
be identified with the high temperature expansion of a Curie--Weiss type law
(per ion and in the same units we are considering in this work), \( \hat{\chi }^{tet}=\frac{s(s+1)}{3T}\left( 1+\frac{\theta }{T}\right)  \),
where \( \theta =2s(s+1)J_{1} \). Therefore, this model reproduces the behavior
predicted by the Curie law at very high temperatures. However, the value of
the Curie--Weiss temperature in this type of lattice is given by \cite{smart.1966}
\begin{equation}
\label{curie.weiss.temp}
\theta ^{CW}=4J_{1}s(s+1)=2\theta .
\end{equation}
 The reason for this deviation is, obviously, the fact that we have not considered
the interaction with the neighboring tetrahedra, which is of the same order
of magnitude that the interactions inside the tetrahedron.

To remedy this situation, we can introduce a tetrahedral mean field (TMF) theory
that takes into account, at least in an approximate way, the interaction with
the neighboring ions outside the tetrahedron. Each ion in the tetrahedron interacts
with \( z=3 \) external ions, in the nearest neighbor interaction approximation.

The effect of this interaction with the nearest neighbors outside the tetrahedron,
in the spirit of the MF theory, can be accounted for by introducing a molecular
field proportional to \( z \) and the magnetization per ion \( m=\frac{M}{4} \).
The constant of proportionality can be put without loss of generality as \( 2J_{eff} \).
In the high temperature limit, the magnetization per ion can be put as \( m=\hat{\chi }^{tet}\, \left( H_{0}+2zJ_{eff}m\right)  \),
from which we obtain the expression of the susceptibility in this tetrahedral
mean field (TMF) model 
\begin{equation}
\chi ^{tmf}=\frac{\hat{\chi }^{tet}}{1-2z\, J_{eff}\, \hat{\chi }^{tet}}.
\end{equation}

The value of the \( J_{eff} \) parameter can be estimated as follows: in the
high temperature limit, \( \chi ^{tmf} \) can be put again as the high temperature
expansion of a Curie--Weiss type law. By equating this expansion with the real
Curie--Weiss law up to \( \frac{1}{T^{2}} \) terms, we reach at the condition
\begin{equation}
\label{condition.high}
\theta +2z\, J_{eff}\frac{s(s+1)}{3}=\theta ^{CW}.
\end{equation}
 Were the interaction with further neighbors negligible with respect to the
first neighbor interactions, relation \eqref{curie.weiss.temp} would be exact
and, therefore, by using the relation between \( \theta  \) and \( J_{1} \)
obtained in the non--interacting tetrahedra case, we would have \( J_{eff}=J_{1} \),
which is completely similar to the value obtained in the standard mean field
theory.

Of course, second and further nearest neighbor interactions are always present
in real systems. Thus, the value of \( \theta ^{CW} \) obtained from experimental
data can contain additional contributions coming from next nearest neighbors
and so on. For this reason, relation \eqref{curie.weiss.temp} is only an approximate
one, and so is \eqref{condition.high}. In order to compare the predictions of
the model with experimental measurements, it is preferable to consider the interaction
inside the tetrahedron, \( J_{1} \), and the interaction with nearest neighbors
outside the tetrahedron, \( J_{eff} \), as adjustable parameters. The difference
between \( J_{1} \) and \( J_{eff} \) provides a way of estimating the value
of additional interactions not explicitly accounted for in this model. In fact,
interactions with further neighbors can be explicitly accounted for in this
model in the following way: let us consider the interaction with \( z' \) next
nearest neighbors (9 for the pyrochlore lattice) in the mean field approximation.
In this case, the magnetization per ion will be given by \( m=\hat{\chi }^{tet}\, \left( H_{0}+2zJ_{1}m+2z'\, J_{2}m\right)  \),
where \( J_{2} \) represents the coupling with next nearest neighbors, and
the magnetic susceptibility is given by
\begin{equation}
\label{suscep.tmf}
\chi ^{tmf}=\frac{\hat{\chi }^{tet}(J_{1},T)}{1-6(J_{1}+3J_{2})\, \hat{\chi }^{tet}(J_{1},T)},
\end{equation}
 where we have made use of the fact that for the pyrochlore lattice \( z=3 \)
and \( z'=9 \) and that the difference between \( J_{1} \) and \( J_{eff} \)
comes from these additional interactions, that is, \( J_{eff}=J_{1}+3J_{2} \).

\textit{Comparison with experimental data.-} In order to test the validity of
the above model, it is important to compare its predictions with magnetic susceptibility
data. We have selected the following systems whose magnetic ions form a pyrochlore
lattice: ZnCr\( _{2} \)O\( _{4} \) \cite{huber.2000}, Gd\( _{2} \)Ti\( _{2} \)O\( _{7} \)
\cite{raju.1999}, Y\( _{2} \)Mo\( _{2} \)O\( _{7} \) \cite{greedan.1986},
and CdFe\( _{2} \)O\( _{4} \) \cite{Ostorero.1989}, where the value of the
magnetic ion spin is \( s=\frac{3}{2} \), \( s=\frac{7}{2} \), \( s=1 \),
and \( s=\frac{5}{2} \), respectively.
\begin{figure*}
\includegraphics[width=\textwidth]{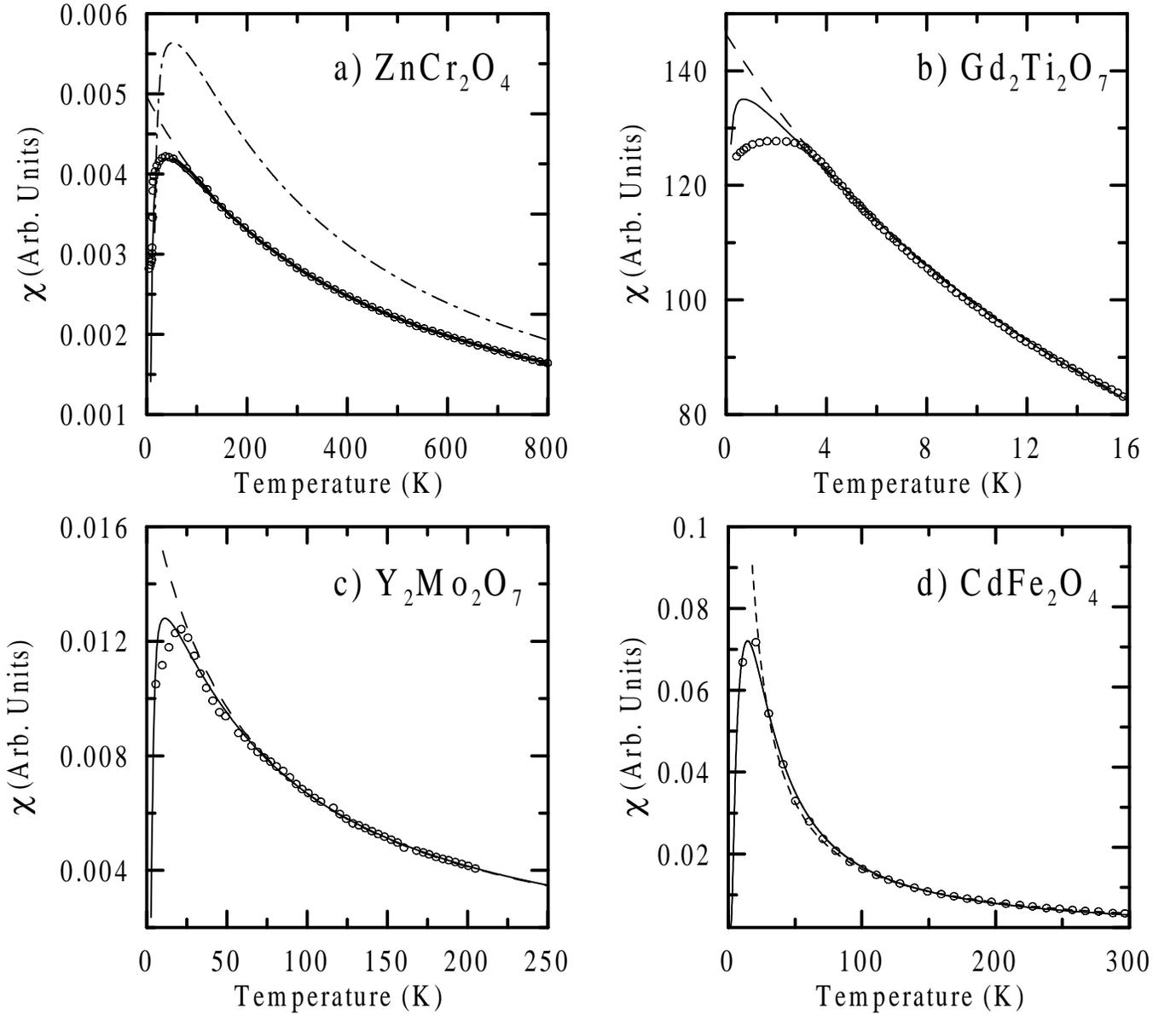}
\caption{\label{fig.data}Comparison with experimental susceptibility data of (a) ZnCr\protect\( _{2}\protect \)O\protect\( _{4}\protect \),
(b) Gd\protect\( _{2}\protect \)Ti\protect\( _{2}\protect \)O\protect\( _{7}\protect \),
(c) Y\protect\( _{2}\protect \)Mo\protect\( _{2}\protect \)O\protect\( _{7}\protect \),
and (d) CdFe\protect\( _{2}\protect \)O\protect\( _{4}\protect \). The open
circles represent experimental data, whereas the solid lines are fits to expression
\eqref{suscep.tmf}. The dashed lines represent fits to the Curie--Weiss law.
The dot--dashed line in the case of ZnCr\protect\( _{2}\protect \)O\protect\( _{4}\protect \)
represents the prediction of the isolated tetrahedra model with the parameters
obtained from the fit to the Curie--Weiss law.}
\end{figure*}

The temperature dependence of the magnetic susceptibility in these materials
can be seen in fig. \ref{fig.data}. All of them follow a Curie--Weiss law from
high temperatures to well below the Curie--Weiss temperature. The predictions
of our TMF theory are also plotted in fig. \ref{fig.data}. The values of \( J_{1} \)
and \( J_{2} \), which we obtain by fitting the data points above \( T_{c} \),
have been quoted in table \ref{tab.parameters}. Here \( T_{c} \) marks the
transition to the low temperature phase, which is either long range order (LRO)
or spin--glass (SG). The data for Gd\( _{2} \)Ti\( _{2} \)O\( _{7} \) are
from AC susceptibility measurements. Also, the predictions of the non--interacting
tetrahedra model have been represented for the case of ZnCr\( _{2} \)O\( _{4} \),
in order to show the deviations of the predictions of this model from the experimental
data. The case of CdFe\( _{2} \)O\( _{4} \) is somewhat special and will be
analyzed below.

As we can see from observation of the figure and the values collected in table
\ref{tab.parameters}, the agreement between theory and experiment is quite
good. In the case of ZnCr\( _{2} \)O\( _{4} \), the agreement is excellent
in the entire temperature range. For the other two systems, there is only qualitative
agreement. Nevertheless, we feel that the agreement between theory and experiment
is good enough to assess the validity of our model, at least, from a qualitative
point of view.

It is important to notice that the values predicted for \( J_{1} \) are systematically
smaller than the values of \( J_{eff}=J_{1}+3J_{2} \), which are close to the
ones obtained from fits to a Curie--Weiss law, which is consistent with the
idea mentioned in the previous section about additional interactions from further
neighbors.
\begin{table*}[h!]
\begin{tabular}{|c|r|r|r|r|r|r|r|}
\hline 
\multicolumn{1}{|c|}{Material}&
\multicolumn{1}{|c|}{\( \theta ^{CW} \)}&
\multicolumn{1}{|c|}{\( J^{CW} \) \footnote{Value of the antiferromagnetic exchange coupling obtained from the equation
\( \theta ^{CW}=4J^{CW}s(s+1) \)}}&
\multicolumn{1}{c|}{ \( T_{m}^{exp} \) \footnote{Experimental value of the position of the maximum}}&
\multicolumn{1}{c|}{\( T_{c} \)}&
\multicolumn{1}{|c|}{\( J_{1} \)}&
\multicolumn{1}{c|}{\( J_{2} \)}&
 \( T_{m}^{fit} \) \footnote{Calculated from eq. \eqref{maxima}}\\
\hline 
\multicolumn{1}{|l|}{ZnCr\( _{2} \)O\( _{4} \)}&
-388&
 -25.9&
37.4&
12 (LRO)&
 -19.2&
-2.3&
38.0\\
\hline 
\multicolumn{1}{|l|}{Gd\( _{2} \)Ti\( _{2} \)O\( _{7} \)}&
-21&
 -0.33&
2.0&
0.97 (LRO)&
 -0.20&
-0.05&
0.7\\
\hline 
\multicolumn{1}{|l|}{Y\( _{2} \)Mo\( _{2} \)O\( _{7} \)}&
-61&
 -7.625&
\multicolumn{1}{c|}{--}&
18 (SG)&
 -7.13&
 -0.02&
11.2\\
\hline 
\multicolumn{1}{|l|}{CdFe\( _{2} \)O\( _{4} \)}&
0&
0&
\multicolumn{1}{r|}{17.0}&
10 (LRO)&
-5.23&
4.1&
14.4\\
\hline 
\end{tabular}

\caption{\label{tab.parameters}Parameters of the TMF theory. All of them are quoted
in K.}
\end{table*}

An additional prediction of our model is the existence of a maximum appearing
in the \( \chi  \) versus \( T \) plots. The position of this maximum, \( T_{m} \),
presented in table \ref{tab.parameters}, can only be calculated numerically,
as its determination involves solving a transcendental equation. However, it
is found that it follows this empirical law 
\begin{equation}
\label{maxima}
T_{m}=-\frac{(s+1)\, J_{1}}{1.273}.
\end{equation}
 It is important to notice that the position of the maximum is determined only
by the value of the interaction \emph{inside} the tetrahedron. We see that the
value of \( T_{m} \) predicted by the model in the case of ZnCr\( _{2} \)O\( _{4} \)
is very close to the experimental value. However, in the case of Gd\( _{2} \)Ti\( _{2} \)O\( _{7} \)
the predicted maximum occurs at \( \sim 1/2 \) of the experimental value. In
the case of Y\( _{2} \)Mo\( _{2} \)O\( _{7} \) there is no maximum in the
paramagnetic phase as there is a transition to a SG at around 18 K.

As commented above, the case of CdFe\( _{2} \)O\( _{4} \) is especial, because
in this material, next nearest neighbor interactions are of the same order of
magnitude than the nearest neighbor ones, but ferromagnetic, which leads to
an experimental value of the Curie--Weiss temperature equal to zero. However,
the present model is able to describe even this limiting case, and the values
of the various couplings obtained from the fit are very reasonable. Moreover,
the position of the maximum predicted by the model is in very good agreement
with the experimental value, and inside the experimental uncertainty.

\textit{Conclusions.-} In this work we have analyzed the temperature dependence
of the magnetic susceptibility data of an arrangement of magnetic ions of spin
\( s \) in a pyrochlore lattice. To do this, we have developed a tetrahedral
mean field theory taking as a starting point the exact susceptibility of a set
of four interacting spins in the corners of an isolated tetrahedron. For the
sake of completeness, we have also briefly analyzed the susceptibility predicted
by a non--interacting tetrahedra model. We reach the conclusion that, even though
the non--interacting model provides an adequate description of the high temperature
region, it fails to quantitatively describe the temperature dependence of the
susceptibility at lower temperatures, as the \( \frac{1}{T^{2}} \) term in
this model is \( \frac{1}{2} \) of the actual value. However, once we incorporate
the interactions with the rest of nearest neighbors in terms of a mean field,
we find a quite good agreement between the theoretical predictions and experimental
data. Especially important is the prediction of the appearance of a maximum
in \( \chi  \) versus \( T \). The position of the maximum has been calculated,
and a reasonable agreement is found. At this point, it is very difficult to
say if the deviations are an intrinsic problem of the model. To clarify this
point, more good quality experimental data in a variety of systems whose susceptibility
maxima lie well above \( T_{c} \) would be necessary.

One feature of this model is that it can be very easily generalized to include
additional interactions coming from further neighbors and more exotic systems
where, for example, the interaction with the nearest neighbors is antiferromagnetic
and is ferromagnetic with next nearest neighbors, as we have done for the case
of CdFe\( _{2} \)O\( _{4} \), finding a very good agreement between theory
and experiment.

Of course, this work is not conclusive, in the sense that it does not solve
the problem of the anomalous behaviors found in these types of systems. It does
not provide an explanation of why some of these materials exhibit a long range
order state at very low temperatures, whereas other are in a spin liquid state.
Moreover, nothing has been said about the disorder always present in any material,
which has been suggested to be related to the appearance of behaviors characteristic
of spin glasses. These subjects are out of the scope of this paper, though they
should constitute the direction of future work. 

In any case, we think that the present model provides a good starting point
for such investigations.

\begin{acknowledgments}

Angel García Adeva wants to acknowledge the Spanish MEC for financial support
under the Subprograma General de Formación de Personal Investigador en el Extranjero.

David L. Huber wishes to thank S. Oseroff and C. Rettori for stimulating his interest in the properties of pyrochlore antiferromagnets.

\end{acknowledgments}

\end{document}